\documentclass[twocolumn,aps,pra,preprintnumbers,bibnotes10pt,superscriptaddress]{revtex4}

\usepackage{graphicx}
\usepackage{dcolumn}
\usepackage{epsfig}
\usepackage{bm}
\usepackage{array}
\usepackage{amsmath}
\usepackage{soul}

\usepackage{graphicx,subfigure,wrapfig}
\usepackage{float}
\usepackage{color}
\usepackage{epstopdf}
\usepackage{amsmath}
\usepackage{braket}
\usepackage{amsthm}
\usepackage{amssymb}
\usepackage{amsfonts}
\usepackage{graphicx,subfigure}
\usepackage{txfonts}
\usepackage{ulem}
\usepackage{mathtools}
\usepackage{bm}
\usepackage{bbm}
\usepackage{hyperref}
\usepackage[hyphenbreaks]{breakurl} 
\usepackage{natbib}
\usepackage{comment}

\usepackage{siunitx}
\usepackage{ulem}
\usepackage{bbm}
\usepackage{color}
\usepackage{soul}
\usepackage[export]{adjustbox}
\usepackage{hyperref}
\usepackage{url}
\usepackage{breakurl}    
\definecolor{mygreen}{rgb}{0,0.5,0}
\definecolor{mygrey}{rgb}{0.5,0.5,0.5}
\definecolor{myred}{rgb}{0.75,0,0}
\definecolor{myblue}{rgb}{0,0,0.75}
\definecolor{mymagenta}{cmyk}{0,1,0,0.12}
\definecolor{mycyan}{cmyk}{1,0,0,0.12}
\definecolor{myorange}{rgb}{1,0.5,0}
\definecolor{myviolet}{rgb}{0.5,0.0,0.75}

\newcommand{\btext}[1]{{\color{myblue}#1}}

\definecolor{mybrown}{rgb}{0.542969,0.269531, 0.0742188} 

\newcommand{\commentout}[1]{}

\newcommand{\vect}[1]{\bm{#1}}

\begin{document}

\title{Mach-Zehnder atom interferometry with non-interacting trapped Bose Einstein condensates}
%
%

\author{T. Petrucciani}
\altaffiliation{These authors contributed equally to this work}
\affiliation{Istituto Nazionale di Ottica, Consiglio Nazionale delle Ricerche (CNR-INO), Largo Enrico Fermi 6, 50125 Firenze, Italy} 

\author{A. Santoni}
\altaffiliation{These authors contributed equally to this work}
\affiliation{University of Naples “Federico II”, Via Cinthia 21, 80126 Napoli, Italy} 
\affiliation{European Laboratory for Nonlinear Spectroscopy (LENS), Via N. Carrara 1, 50019 Sesto Fiorentino, Italy}

\author{C. Mazzinghi} 
\affiliation{Istituto Nazionale di Ottica, Consiglio Nazionale delle Ricerche (CNR-INO), Largo Enrico Fermi 6, 50125 Firenze, Italy} 
\affiliation{European Laboratory for Nonlinear Spectroscopy (LENS), Via N. Carrara 1, 50019 Sesto Fiorentino, Italy}

\author{D. Trypogeorgos}
\affiliation{Institute of Nanotechnology, Consiglio Nazionale delle Ricerche (CNR-Nanotech), via Monteroni 165, 73100, Lecce, Italy}

\author{F. S. Cataliotti}
\affiliation{Istituto Nazionale di Ottica, Consiglio Nazionale delle Ricerche (CNR-INO), Largo Enrico Fermi 6, 50125 Firenze, Italy} 
\affiliation{European Laboratory for Nonlinear Spectroscopy (LENS), Via N. Carrara 1, 50019 Sesto Fiorentino, Italy}

\author{M. Inguscio}
\affiliation{Istituto Nazionale di Ottica, Consiglio Nazionale delle Ricerche (CNR-INO), Largo Enrico Fermi 6, 50125 Firenze, Italy} 
\affiliation{European Laboratory for Nonlinear Spectroscopy (LENS), Via N. Carrara 1, 50019 Sesto Fiorentino, Italy}
\affiliation{University of Florence, Physics Department, Via Sansone 1, 50019 Sesto Fiorentino, Italy}

\author{G. Modugno}
\affiliation{European Laboratory for Nonlinear Spectroscopy (LENS), Via N. Carrara 1, 50019 Sesto Fiorentino, Italy}
\affiliation{University of Florence, Physics Department, Via Sansone 1, 50019 Sesto Fiorentino, Italy}

\author{A. Smerzi}
\affiliation{Istituto Nazionale di Ottica, Consiglio Nazionale delle Ricerche (CNR-INO), Largo Enrico Fermi 6, 50125 Firenze, Italy} 
\affiliation{European Laboratory for Nonlinear Spectroscopy (LENS), Via N. Carrara 1, 50019 Sesto Fiorentino, Italy}
\affiliation{QSTAR, Largo Enrico Fermi 2, 50125 Firenze, Italy}

\author{L. Pezz\`e}
\affiliation{Istituto Nazionale di Ottica, Consiglio Nazionale delle Ricerche (CNR-INO), Largo Enrico Fermi 6, 50125 Firenze, Italy} 
\affiliation{European Laboratory for Nonlinear Spectroscopy (LENS), Via N. Carrara 1, 50019 Sesto Fiorentino, Italy}
\affiliation{QSTAR, Largo Enrico Fermi 2, 50125 Firenze, Italy}

\author{M. Fattori}
\affiliation{Istituto Nazionale di Ottica, Consiglio Nazionale delle Ricerche (CNR-INO), Largo Enrico Fermi 6, 50125 Firenze, Italy} 
\affiliation{European Laboratory for Nonlinear Spectroscopy (LENS), Via N. Carrara 1, 50019 Sesto Fiorentino, Italy}
\affiliation{University of Florence, Physics Department, Via Sansone 1, 50019 Sesto Fiorentino, Italy}

\date{\today}

\begin{abstract}
The coherent manipulation of a quantum wave is at the core of quantum sensing. 
For instance, atom interferometers require linear splitting and recombination processes to map the accumulated phase shift into a measurable population signal. 
Although Bose Einstein condensates (BECs) are the archetype of coherent matter waves, their manipulation between trapped spatial modes has been limited by the strong interparticle collisions.
Here, we overcome this problem by using BECs with tunable interaction trapped in
an innovative array of double-well potentials and exploiting quantum tunneling to realize linear beam splitting. 
We operate several Mach-Zehnder interferometers in parallel, canceling common-mode potential instabilities by a differential analysis, thus demonstrating a trapped-atom gradiometer.  
Furthermore, by applying a spin-echo protocol, we suppress additional decoherence sources and approach unprecedented coherence times of one second.
Our interferometer will find applications in precision measurements of forces with a high spatial resolution and in linear manipulation of quantum entangled states for sensing with sub shot-noise sensitivity. 
\end{abstract}
\maketitle



Bose-Einstein Condensates (BECs) of ultracold atomic gases~\cite{KetterleRMP2002, CornellRMP2002} are recognized as powerful tools for both fundamental research~\cite{KetterleBEC, LeggettRMP2001} and emerging quantum technologies~\cite{Bloch, PezzeRMP2018, AmicoRMP2022}.
Often regarded as the matter-wave analog of lasers  \cite{KetterleAL,BlochAL}, BECs hold particular promise in the domain of high precision measurements~\cite{Cronin}.
Thanks to their remarkably low momentum spread \cite{Herbst2024-hj}, they are currently exploited in free-falling atom interferometers within drop towers~\cite{Kasevich, Abe_2021, PhysRevLett.110.093602, CondonPRL2019, PhysRevLett.127.100401, Beaufils2022-sp, Zahan} and earth-orbiting research laboratories~\cite{cal}.
Trapped atom interferometers~\cite{Ferrari, PhysRevA.85.013639, PhysRevA.91.053616, mueller} that exploit the unique properties of BECs \cite{Shin, Schumm} promise to bring measurements of gravity, inertial forces and electromagnetic fields into a new realm. The large spatial coherence of a quantum degenerate gas enables splitting a BEC in two interferometric modes with large separation, allowing for high-sensitive measurements. Moreover, the quantum-limited size of a BEC provides Heisenberg-limited spatial resolution for sensing applications.

However, the full development of trapped BEC interferometry has been fundamentally hindered by strong interparticle interactions. 
Collisions between atoms induce decoherence of the interferometric signal~\cite{PhysRevLett.77.3489, JavanainenPRL1997} and obstruct the realization of linear beam splitters~\cite{PezzePRA2006, IMZ}. 
Previous valuable attempts to overcome these issues \cite{BerradaNATCOMM2013,BerradaPRA2016} have been limited by the short coherence times and a reduction of the interference contrast. 
Additionally, interatomic collisions prevent the linear manipulation of quantum entangled states, which is necessary for operating the interferometer below the shot-noise limit \cite{EsteveNATURE2008}. 
A possible solution involves tuning the collisional scattering length to zero via magnetic Feshbach resonances -- as already employed to suppress interaction-induced decoherence in Bloch oscillation~\cite{Gustavsson, Fattori}. However, non-interacting gases make the interferometer extremely sensitive to imperfections in the trapping potential \cite{Balestri, PhysRevLett.117.275301}.
Therefore, despite early optimistic expectations, atom interferometry based on BECs confined in two distinct spatial modes has yet to fully realize its potential in the field of precision metrology.

\begin{figure}[t!]
    \centering
    \includegraphics[width=\columnwidth, trim=3cm 0cm 3cm 0cm, clip]{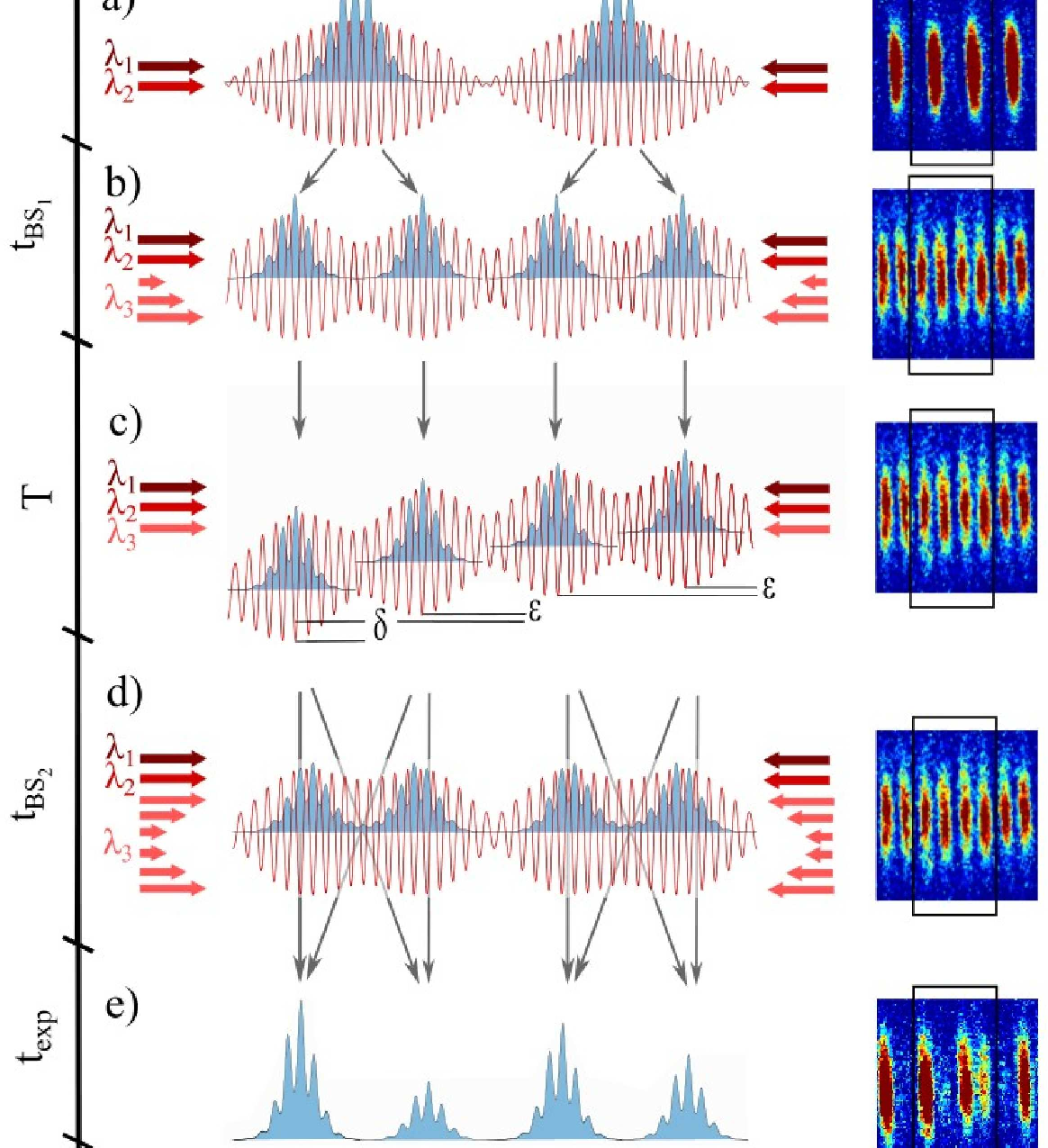}
    \caption{{\bf Gradiometer with trapped BEC Mach-Zehnder interferometers.} 
    The trapping potential (red line) is realized with two Beat-Note Superlattices (BNSL)~\cite{PhysRevLett.127.020601} formed by three retro-reflected lasers with wavelengths $\lambda_1$, $\lambda_2$ and $\lambda_3$: the intensity and manipulation of each lattice is schematically illustrated by the lateral arrows. 
    Superposed to the trapping potential, we show 1D integrated profiles of the BEC density, while experimental absorption images are reported on the right. 
    Different sections correspond to different stages of gradiometer operation implemented with non-interacting atoms. 
    a) \textit{BEC array.} The two optical lattices $\lambda_{1}$ and $\lambda_2$ form a BNSL, with effective lattice sites separated by $\sim$\SI{10}{\micro\meter} and loaded with independent BECs.
    The lattice depth is enough to suppress the tunneling between sites. 
    b) \textit{Raising barrier beam-splitter.} The amplitude of the third lattice, with wavelength $\lambda_3$, is raised in a time $t_{\rm BS_1}$ splitting the clouds in two spatial modes separated by $d=$\SI{5}{\micro\meter}. 
    c) \textit{Phase shift}. 
    During the interrogation time $T$, an external homogeneous force induces an equal energy shift $\varepsilon$ in both DWs.
    Instead, a spatial dependent force introduces a differential energy shift $\delta$. 
    d) \textit{Tunneling linear beam-splitter.} The third lattice amplitude is first lowered and then raised again in a time $t_{\rm BS_2}$ to achieve a 50\% tunneling probability between the two modes.
    e) \textit{Detection.} Final populations in the two modes of the interferometers are measured via standard absorption imaging.
    }
    \label{fig 1}
\end{figure}

In this work we overcome current roadblocks and demonstrate the long sought operation of a full Mach-Zehnder Interferometer (MZI) with non-interacting BECs trapped in two spatially separated modes.    
The key ingredient is an innovative potential made of an array of Double-Well (DW) traps where several interferometers operate simultaneously. 
In this way, residual instabilities of the trap, that could mask the coherence of a single sensor, are common mode and can be canceled using a differential analysis. 
Our device realizes the first trapped atom gradiometer reported in the literature~\cite{NotaKasevich}. 
Each MZI (see the scheme in Fig.~\ref{fig 1}) consists of a beam splitter, an interrogation time, and a final beam splitter that maps the accumulated relative phase into a population imbalance.
The beam splitters operate in the linear regime with near-unit contrast by canceling inter-particle interactions via a magnetic Feshbach resonance and by tuning the height of the barrier separating each DW - changing the tunneling energy.
By implementing a spin-echo protocol \cite{PhysRev.80.580}, we suppress residual technical noise and extend the interferometer’s coherence time to approximately 1 second — nearly two orders of magnitude longer than previously reported values \cite{BerradaNATCOMM2013, BerradaPRA2016}.

The experimental system consists of a BEC of $^{39}$K atoms with tunable interactions~\cite{PhysRevA.86.033421}, manipulated by an innovative array of DWs \cite{Petrucciani}.
During the final part of the evaporation stage, two collinear optical lattices aligned along the horizontal $x$ directions are switched on adiabatically. 
Their wavelengths $\lambda_1 \simeq1013$ nm and $\lambda_2\simeq 1064$ nm create a Beat-Note Superlattice (BNSL)~\cite{PhysRevLett.127.020601} with a spacing equal to $\tfrac{1}{2}\tfrac{\lambda_1 \lambda_2}{\lambda_2-\lambda_1} \sim 10 \ \mu$m. 
In this way, several independent BECs can be loaded in the minima of the potential, Fig.~\ref{fig 1}a.
By using a third optical lattice superimposed to the others, and with wavelength $\lambda_3\simeq1120$ nm, we create an additional effective lattice with spatial periodicity $d=\tfrac{1}{2}\tfrac{\lambda_3 \lambda_1}{\lambda_2-\lambda_1} \sim 5 \ \mu$m.
By raising this optical potential in $t_{\rm BS_1}=10$ ms in the center of each BEC, we form an array of DWs ~\cite{supp} and realize the first beam splitter of each MZI, see Fig.~\ref{fig 1}b.
This operation is followed by an interrogation time $T$, see Fig.~\ref{fig 1}c.
An external homogeneous force provides a linear potential energy that affects each interferometer with an equal energy mismatch $\varepsilon$ between the two wells.
A spatial-dependent force, instead, provides an additional non-common energy difference $\delta$ between the two modes of each MZI, see Fig.~\ref{fig 1}c. 
By performing a further beam splitter (Fig.~\ref{fig 1}d), we map the accumulated phase onto a population difference (Fig.~\ref{fig 1}e).
The energy difference $\delta$ can be then estimated by a differential analysis, as illustrated below. 

\begin{figure}[t!]
    \centering
     \includegraphics[width=0.5\textwidth]{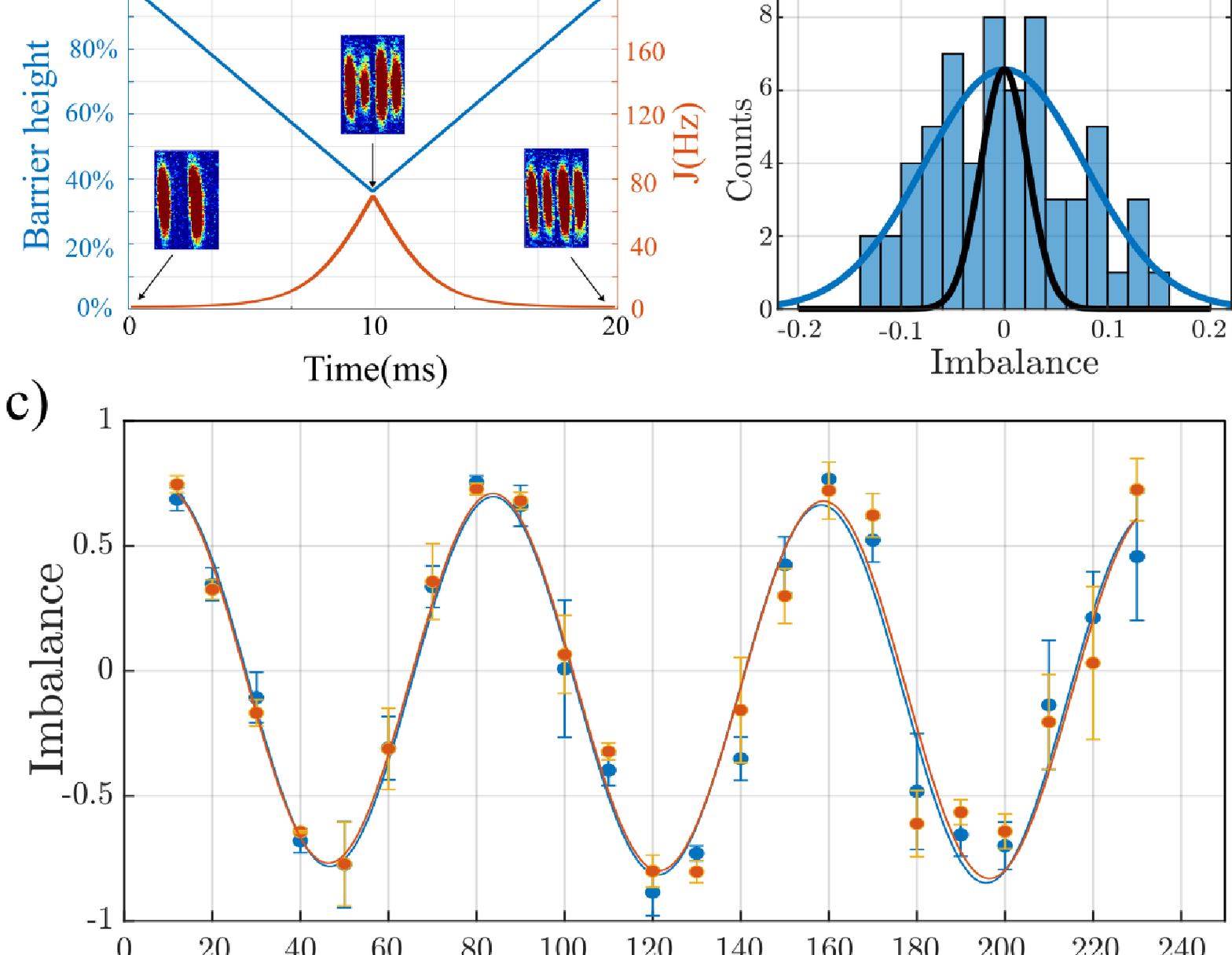}
    \caption{{\bf Tunneling atom beam-splitter.}
    a) Evolution of the DW parameters, barrier height (blue line) and tunneling energy (red line), that allows the BECs initially loaded in the left mode of each DW to tunnel to the right mode with 50$\%$ probability.
    b)  Histogram of the imbalance $z$ for a single DW (see text) after the beam-splitting sequence. The blue line is the Gaussian fit of the data and the black line represents the projection noise. We obtain a noise of 0.078(19) from the width of the Gaussian fit, that is three times larger with respect to the projection noise that is $\sqrt{2/N}\simeq0.023$.
    c) Synchronous Rabi oscillations show equal tunneling energy for two neighboring DWs. The points and the error bars show the mean and  plus/minus one mean standard deviation on 3 measurements. The fitted Rabi frequencies are  \SI[separate-uncertainty = false]{13.4(3)}{\Hz} (blue line) and \SI[separate-uncertainty = false]{13.3(3)}{\Hz} (orange line). }
    \label{fig 2}
\end{figure}

\begin{figure*}[ht!]
    \centering \includegraphics[width=0.9\textwidth]{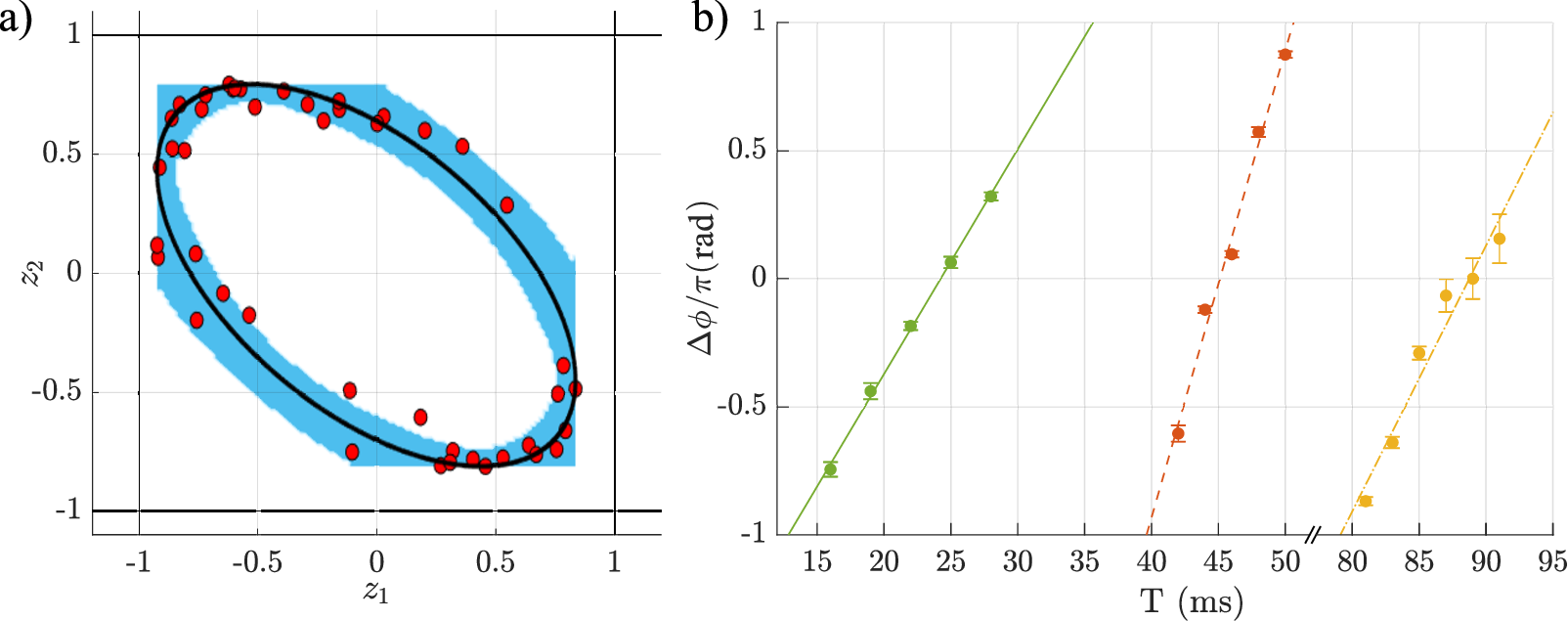}
    \caption{
    {\bf Gradiometric measurements.}
    (a)	Output population imbalances (red dots) of two neighboring DW interferometers. The black line represents the elliptical fit obtained using a maximum likelihood method (see text and Methods). The blue shaded region represents 90\%  confidence area of the fit. 
    The estimated center of the ellipse is C$_1$=-0.0387(8), C$_2$=-0.066(5); the estimated visibilities are V$_1$=0.89(2), V$_2$=0.85(2).
    (b)	$\Delta \phi$ as a function of $T$ in three different time intervals around 20 ms, 45 ms and 85 ms. 
    The three set of measurements allow to estimate the trap frequency values from a linear fit: \SI[separate-uncertainty = false]{17.9(6)}{\Hz} (solid line), \SI[separate-uncertainty = false]{21.7(9)}{\Hz} (dashed line), \SI[separate-uncertainty = false]{16.4(3)}{\Hz} (line-point). 
    The error bars represent the standard deviation of the mean value computed with the Bootstrap analysis (see Methods).
    Each point is obtained from the analysis of about 30 $z_1$ vs $z_2$ data.
    The three trap frequency values agree with the ones estimated from the oscillations of trapped BECs, 
    these are \SI[separate-uncertainty = false]{18.5(2)}{\Hz}, \SI[separate-uncertainty = false]{20.5(4)}{\Hz}, \SI[separate-uncertainty = false]{16.8(2)}{\Hz}. The different values are due daily variations of the vertical beam, used to create the harmonic potential.
    }
    \label{fig 3}
\end{figure*}

The second beam-splitter exploits the linear evolution of non-interacting atoms tunneling through the central barrier of each DW.
We achieve a balanced beam splitter by decreasing the barrier height and then raising the barrier back to its initial value, see Fig.~\ref{fig 2}a.
This is done by acting on the third lattice depth leaving unchanged the first two. 
In Fig.~\ref{fig 2}b, we show the distribution of the population imbalance $z_j=(N_{L,j}-N_{R,j})/(N_{L,j}+N_{R,j})$ after the splitting sequence (see histogram in Fig.~\ref{fig 2}b) starting with all the atoms in the left mode. 
Here, $N_{R,j}$ and $N_{L,j}$ are the number of atoms in the right and left side of the $j$-th DW ($j=1,2$).
Residual instabilities of the energy mismatch between the two modes prevents the achievement of shot noise limited splitting (black line in Fig.~\ref{fig 2}b).
Shorter sequences with larger tunneling energies could help to approach such limit, but they could cause unwanted excitations of the atoms to higher lying modes.  
The possibility to operate simultaneously two or more interferometers requires that the tunneling energy of neighboring DWs is equal. 
This is demonstrated in Fig.~\ref{fig 2}c, where we plot the population imbalance $z_j$ (for two neighboring DWs) as a function of tunneling time for fixed barrier height.
The results show near-unity visibility of Rabi oscillations that would be prevented by particle-particle interactions~\cite{PezzePRA2006}.

In the following, we select the two central DWs of the BNSL, each loaded with $N \approx 3000$ atoms. An external harmonic confinement of frequency $\omega / 2\pi$, created by an additional vertical laser beam, causes a differential energy mismatch $\delta = m \omega^2 d^2$, where $m$ is the atomic mass.
This provides a phase difference between two interferometers equal to $\Delta \phi = \phi_1 - \phi_2 = \frac{\delta}{\hbar} T$, where $\phi_j$ is the interferometric phase acquired by the the $j$-th interferometer.
A plot of the two population imbalances, $z_1$ versus $z_2$, is shown in Fig.~\ref{fig 3}a.  
For two perfectly correlated interferometers with sinusoidal signals $z_j = V_j \sin(\phi_\varepsilon+\phi_j)+C_j$, where $V_j$ are visibilities, $C_j$ are oscillation offsets \btext{\cite{comment}} and $\phi_\varepsilon$ is a common-mode phase noise uniformly distributed between $0$ and $2\pi$, the measurements scatter along an ellipse whose eccentricity depends on the relative phase $\Delta \phi$~\cite{FosterOPTLETT2002, FixlerSCIENCE2007, RosiNATURE2014},
\begin{equation}
\tilde{z}_1^2 + \tilde{z}_2^2 -2 \tilde{z}_1 \tilde{z}_2 \cos(\Delta \phi) - \sin^2(\Delta \phi)=0,    
\end{equation}
where $\tilde{z}_j = (z_j-C_j)/V_j$.
Loss of correlations between the two interferometers, mainly due to the interactions and uncontrolled spatial inhomogeneities (see below), causes a spread of the experimental measurements out of the ellipse.
We perform a multiparameter maximum likelihood analysis of the experimental data~\cite{ML} and extract both the differential signal $\Delta \phi$ and the uncorrelated noise $\sigma_{\Delta\phi}$ (see Methods).
To study $\Delta \phi$ as a function of $T$, we collect three sets of measurements of $\Delta\phi$ chosen in a $-\pi +\pi$ interval around $T=20$ ms, 45 ms and 85 ms (collected in three different days), see Fig.~\ref{fig 3}b.
As expected, $\Delta \phi$ increases linearly with $T$, with a slope proportional to $\delta$.
The interaction is set on the zero crossing of a broad Feshbach resonance around 350 Gauss for atoms in the absolute ground state $\ket{F=1,m_F=1}$, where $F$ is the hyperfine atom number \cite{Fattori}.
The interferometric determination of the external trapping frequency 
from such slopes is in agreement, within the error bars, with the value 
determined using atom sloshing in the external harmonic potential (see Fig. \ref{fig 3}b and \cite{supp}).

\begin{figure}[t!]
    \centering
     \includegraphics[width=\columnwidth]{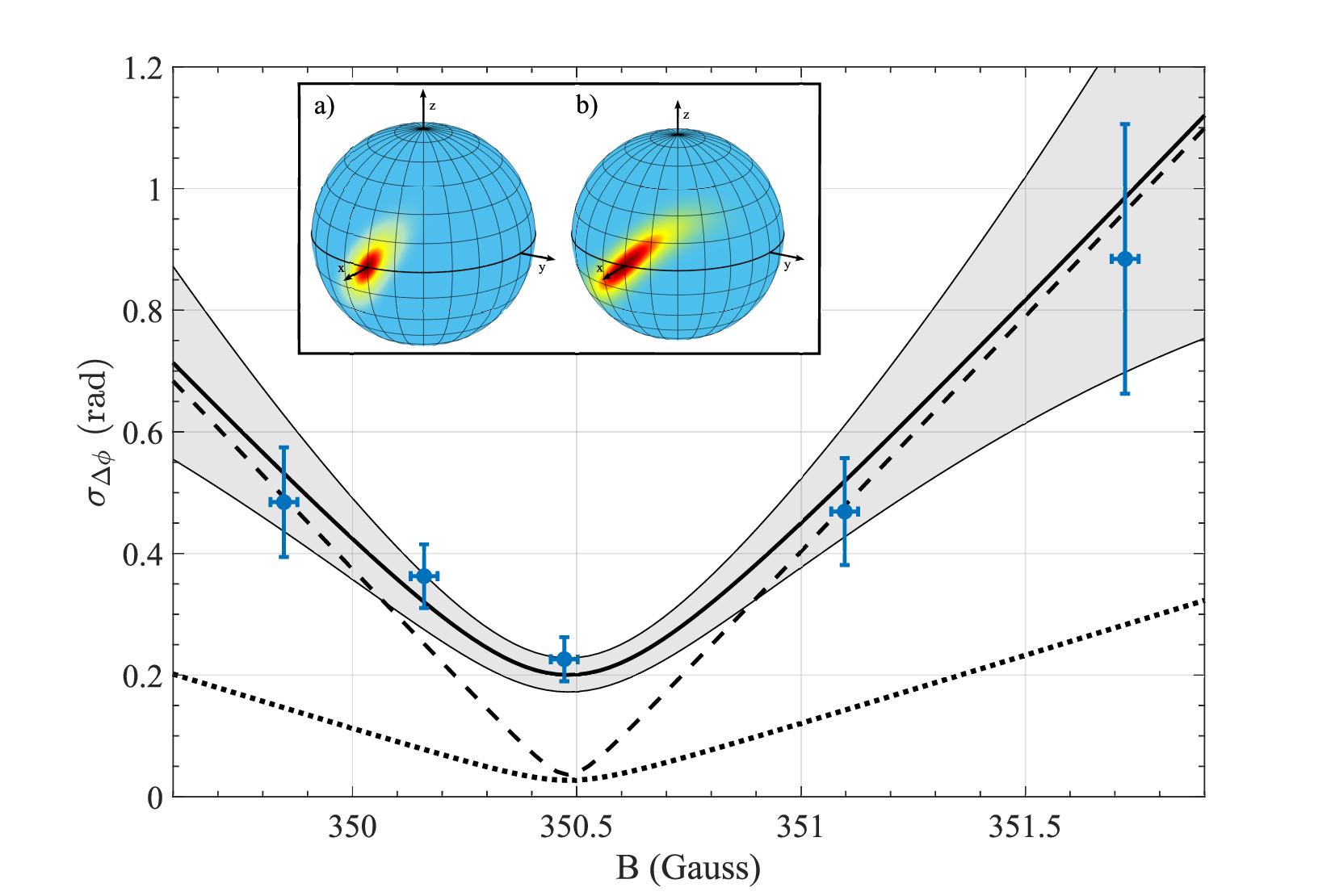}
    \caption{{\bf Interaction-induced decoherence.} 
    Uncorrelated noise $\sigma_{\Delta \phi}$ as a function of the magnetic field $B$ tuning the particle-particle interaction scattering length. 
    Dots are experimental results obtained at $T=70$ ms. 
    The vertical error bars are the standard deviation of the mean value computed with the Bootstrap analysis. The horizontal ones instead are the error on the determination of the magnetic field~\cite{supp}.
    The dotted and dashed lines are Eq.~(\ref{sigmaphi}) for $\sigma_{\rm BS}^2=0$ and $\sigma_{\rm BS}^2=0.004$ (obtained from an independent experimental characterization~\cite{supp}).
    The solid line is the result of numerical simulations that  further include, see Eq.~(\ref{sigmaphi}), an extra stochastic phase noise term $\sigma_{\mathrm{tech}} = 0.15$ rad to fit the experimental data, see~\cite{supp}.
    The gray region show statistical mean squared fluctuations in the maximum likelihood fit due to the finite sample size (about 30 measurements). 
    \textit{Inset}: Wigner distribution of the time-evolved state represented on the generalized Bloch sphere. In a), only quantum fluctuations on the first BS are considered  and the dynamics is only due to one-axis-twisting~\cite{KitagawaPRA1993}, whereas in b) further spread of the state is the result of extra fluctuations $\sigma_{\rm BS}^2=0.004$ after the first beam-splitter. }
    \label{fig 4}
\end{figure}

To study the effect of interaction-induced decoherence,
we fix $T$ and analyze $z_1$ vs $z_2$ as a function of the atomic scattering length by tuning the magnetic field $B$.
The corresponding values of $\sigma_{\Delta \phi}$ are shown in Fig.~\ref{fig 4}~(points). As expected, we observe a minimum of the decoherence around $B_{\rm min}=350.45(6)$ Gauss, in agreement with the previously reported value \cite{PhysRevLett.101.190405}.
We compare the experimental results with the model including the effect of interactions, imperfect splitting and technical dephasing.
We predict~\cite{supp} that the uncorrelated noise for an initial coherent state follows the behavior 
\begin{equation} \label{sigmaphi}
\sigma_{\Delta \phi}^2(B) = \frac{2}{N}(1+\sigma_{\rm BS}^2) + 2N \big(1 + N \sigma_{\rm BS}^2\big)~\bigg( \frac{\chi(B) T }{\hbar}\bigg)^2+\sigma_{\mathrm{tech}}^2.
\end{equation}
Here, $N$ is the number of atoms in each interferometer (assumed equal) and $\chi(B) = \chi_{\rm el}(B) + \chi_{\rm dd}$ is the sum of the elastic particle-particle ($\chi_{\rm el}$) and the dipolar ($\chi_{\rm dd}$) interaction within each well and $\sigma_{\mathrm{tech}}$ is the technical dephasing. 
We have $\chi(B) \propto (B-B_{\rm min})$, where the interplay of the attractive $\chi_{\rm dd}$ and the repulsive $\chi_{\rm el}$ gives a specific value $B_{\rm min}$ such that $\chi=0$. $\sigma_{\rm BS}^2$ quantifies the classical fluctuations of the mean value of $z$ due to uncontrolled energy mismatch between the two DWs after the first beam splitter.
Such fluctuations enhance the spread of the quantum state on the Bloch sphere due to non linear dynamics (see inset in Fig. \ref{fig 4}) that cause a rotation of the Bloch vector with a speed that is proportional to the distance from the equator.
The experimental data are well reproduced by Eq. \ref{sigmaphi}. 

\begin{figure}[t!]
    \centering
    \includegraphics[width=\linewidth, trim=10cm 0.5cm 20cm 0.5cm, clip]{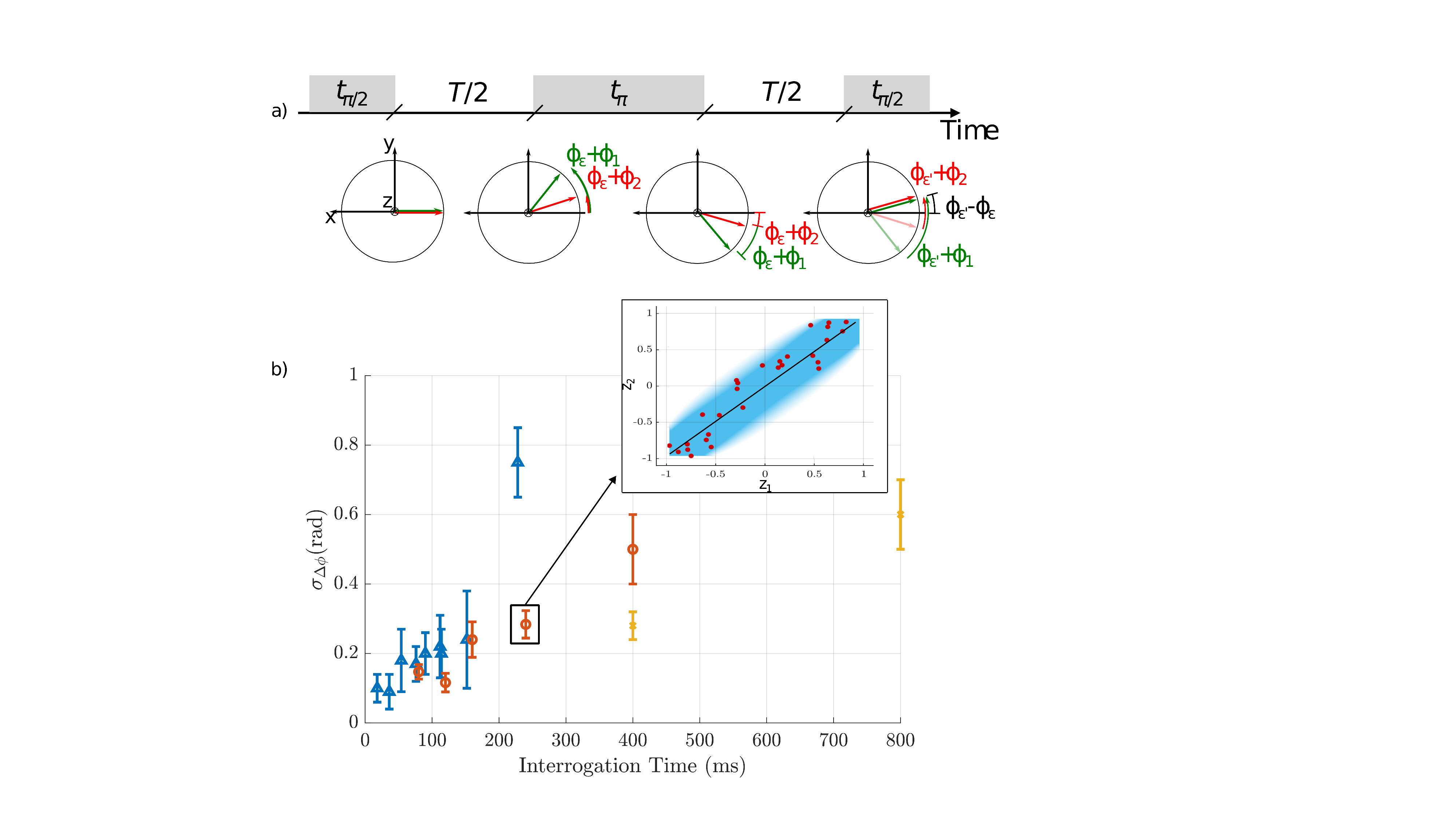}
    \caption{\textbf{Gradiometer with spin-echo.}
    (a) Evolution of the Bloch vectors of two interferometers during the spin-echo sequence. 
    After the first beam-splitter, the arrows point along the $-x$ direction. 
    During the first $T/2$ time interval, the two vectors are rotated by a DC random phase $\phi_{1,2}$ in addition to a common time-varying random phase $\phi_{\varepsilon}$. 
    After a $\pi$ pulse, the vectors are mirrored with respect to the $x$ axis. 
    During the second $T/2$ interval, the random common noise is indicated as $\phi_{\varepsilon'}$. 
    At the end of the sequence, the uncorrelated DC random phase is canceled and perfect correlation between the two vectors is achieved. 
    (b) $\sigma_{\Delta \phi}$ of the gradiometer (blue points) and of the spin echo (red points) as a function of $T$ at $B = 350.45(6)$ G. 
    Yellow points are spin-echo data taken at $B = 350.48(1)$ G. 
    Error bars represent the standard deviation of the mean value computed with the bootstrap method. 
    The inset shows an example of output measurements showing the correlations between $z_1$ and $z_2$. 
    The blue shaded region represents the $90\%$ confidence area of the fit.}
    \label{fig:fig5}
\end{figure}

As shown by the previous analysis, $\sigma_{\mathrm{tech}}$ is the major source of noise that limits the operation of our gradiometer close to $B_{\rm min}$, thanks to our high control of the scattering length. 
Assuming that the technical dephasing is almost constant on the duration of the interferometer, but varying from shot to shot, it can be compensated by a spin-echo protocol. 
This consists of applying an additional $\pi$ pulse at time $T/2$, midway through the interferometric sequence \cite{spin-echo1,spin-echo2}.
The spin echo is realized by a linear ramp of the barrier height, as in Fig. \ref{fig 2}, but reaching a lower value (approximately $30$\% of the initial height). Such sequence lasts half Rabi period and allows to exchange the occupation of the two clouds between the two spatial modes.  
Note that this sequence is achievable due to the unique capability of our apparatus to perform linear rotations around the x-axis of the Bloch sphere.
In Fig. \ref{fig 5}, we compare the measurements of $\sigma_{\Delta \phi}$ as a function of $T$ for both the gradiometer and the spin-echo protocol with $B=B_{\rm min}$. 
Using the gradiometer scheme, we are able to measure a differential phase up to $\approx$ \SI{200}{\milli\second} (blue triangles). 
Instead, with the spin-echo protocol we reach a significant dephasing only after \SI{400}{\milli\second} (red circles). 
At this interrogation time, we further minimize the interaction-induced decoherence by finely tuning the Feshbach magnetic field. 
At the new optimum value $B=350.48(1)$ Gauss, using the spin-echo protocol, we extend the interrogation time up to $T = $\SI{800}{\milli\second} (yellow squares).
The origin of the residual noise is under investigation. 
Multiple $\pi$ pulses can be used to cancel the effect of differential phase drifts occurring on the timescale $T$. 
This will allow to distinguish technical noise sources from fundamental limitations such as finite temperature or residual interparticle interactions.

In conclusion, we report the realization of Mach-Zehnder atom interferometry with trapped BEC, where the linear operations are enabled by the the precise canceling of the scattering length by using a broad magnetic Feshbach resonance. 
Utilizing an innovative array of DWs, we successfully operate two interferometers simultaneously and demonstrate trapped atom gradiometry and spin-echo protocols with synchronous beam splitting pulses. 
Our multiparameter analysis of correlated data provides simultaneous access to both the differential signal and the dephasing -- primarily attributed to residual particle-particle interactions-- while effectively canceling common-mode phase noise. 
Record coherence times of almost one second are reported.



The full control over the spatial mode of a trapped BEC makes our gradiometer ideal for high-precision force measurements with sub-$\mu$m spatial resolution \cite{PhysRevLett.98.063201, PhysRevLett.95.093202, PhysRevA.79.013409, PhysRevLett.133.113403}. 
Correlated analysis of several DW interferometers operating in parallel could be applied to the spatial reconstruction of non-uniform external potentials with common-mode noise rejection.
Additionally, our system is an ideal tool to measure higher-order interaction terms, such as dipolar interactions~\cite{PhysRevLett.101.190405, PhysRevX.12.031018, condmat5020031} and three-body elastic collisions~\cite{PhysRevLett.124.143401}.
Interestingly, our platform offers the unique capability to tune the interaction, enabling the preparation of quantum-entangled states~\cite{PezzePRA2005, PezzeRMP2018,AliceS}, and to cancel the scattering length during the interferometer operations.
This opens the possibility to achieve sub-shot-noise sensitivities, also in the presence of strong environmental dephasing~\cite{Landini, Corgier2023, Eckner2023}.
Finally, long coherence in a trapped BEC interferometer opens interesting perspectives in the development of compact and transportable devices for measurements of inertial forces and navigation applications \cite{JEKELI, BongsNRP2019}. 

\begin{acknowledgments}

We acknowledge financial support by the project SQUEIS of the QuantERA ERA-NET Cofund in Quantum Technologies (Grant Agreement No. 731473 and 101017733) implemented within the European Unions Horizon 2020 Program. We also thank the financial support of the Italian Ministry of Universities and Research under the PRIN2022 project "Quantum sensing and precision measurements with nonclassical states". Finally the project has been co-funded by the European Union - Next Generation EU under the PNRR MUR project PE0000023-NQSTI and under the I-PHOQS 'Integrated Infrastructure Initiative in Photonic and Quantum Sciences'.

\end{acknowledgments}

\section{Methods}

{\bf Beat-note superlattice.}
The array of DWs is realized with three coopropagating lasers at wavelenghts $\lambda_1, \lambda_2$ and $\lambda_3$  retroreflected on a common mirror to form three optical lattices \cite{Petrucciani}. The beam waist of the lasers is $\approx 500 \mu$m and their intensities of 300/380/250 mW provide optical lattice depths of 370/400/240 nK, respectively. Using a polarizing cube we can superimpose the three lattices with an additional dipole trap beam with a wavelength of 1064 nm and with a 12 $\mu$m waist. This laser beam provides a radial confinement of the trapped modes along the $y$ and the vertical $z$ direction of $\approx 180$ Hz. Its weak longitudinal confinement along the 
x direction of $\approx 1$ Hz does not obstacle the simultaneous balancing of many DWs. In addition it is used in combination with another dipole trap beam provided by an intense 100 W IPG laser to perform the final evaporative cooling stage to achieve condensation \cite{PhysRevA.86.033421}. Once a quantum degenerate gas is produced, this second laser is switched off and after a time $\tau$ the intensity of the BNSL with a spacing of 10 $\mu$m is ramped up. By changing the value of $\tau$ we can control the number of DWs populated. We can achieve a maximum number of 8 DWs with $\approx 2 \cdot 10^3$ atoms each.    
The three optical lattice lasers are frequency locked to the same 10 cm length optical cavity with a finesse of $\approx 2000$. In this way cavity length drifts, to the first order, causes only spatial translation of the DWs array.

{\bf Theoretical modeling.}
The single interferometer can be described within a standard two-mode approximation of the many-body order parameter: $\Psi^\dag(\vect{r}) = \psi_{\rm L}(\vect{r}) \hat{c}^\dag_{\rm L} + \psi_{\rm R}(\vect{r}) \hat{c}^\dag_{\rm R}$, where $\psi_{\rm L,R}(\vect{r})$ are the wave function localized in the right and left well, respectively, and $\hat{c}^\dag_{\rm L,R}$ ($\hat{c}_{\rm L,R}$) are bosonic mode creation (annihilation) operators, see e.g. Ref.~\cite{PezzePRA2006} for details.
The mode wave functions are computed as $\psi_{\rm L,R}(\vect{r}) =[\psi_{\rm gs}(\vect{r}) \pm \psi_{\rm ex}(\vect{r})]/\sqrt{2}$,
where $\psi_{\rm gs}(\vect{r})$ are the ground and first excited state of the non-interacting gas of energy $E_{\rm gs}$ and $E_{\rm ex}$, respectively.
We write the interaction potential as $U(\vect{r}-\vect{r}') = g\delta(\vect{r}-\vect{r}') + C_{dd} U_{dd}(\vect{r}-\vect{r'})$, where the first term accounts for s-wave scattering contact interaction, with $g = 4 \pi \hbar^2 a/m$, and the second term describe dipole interaction $U_{dd}(\vect{r}-\vect{r'}) =
\frac{1}{4 \pi} \frac{1-3 \cos \theta}{\vert \vect{r} - \vect{r}'\vert^3}$.
By neglecting the dipolar interaction between nearby wells of the BNSL, we write the interaction contribution to the full Hamiltonian, $H_{\rm int} = \tfrac{1}{2} \int d\vect{r} \int d\vect{r}'~ \Psi^{\dag}(\vect{r}) \Psi^{\dag}(\vect{r}') U(\vect{r}-\vect{r}') \Psi(\vect{r})\Psi(\vect{r}') = (\chi_{\rm cont} + \chi_{\rm dip}) J_z^2$, where $J_z = (\hat{c}^\dag_{\rm L} \hat{c}^\dag_{\rm L} - \hat{c}^\dag_{\rm R} \hat{c}^\dag_{\rm R})/2$.
For the parameters of our experiments, we estimate $\chi_{\rm cont} = 1.4464 \tfrac{a}{x_0}  \sqrt{\omega_y \omega_z} = 0.072 \frac{a}{a_0} ~ {\rm sec}^{-1}$,
where $x_0=1$ $\mu$m is the length unit in the code, $a_0$ is the Bohr radius ($a_0/x_0 =5.3 \times 10^{-5}$), and $\chi_{dd} = -0.01~\frac{C_{dd}}{\hbar x_0^3} = 0.01~{\rm sec}^{-1}$.
To summarize:
\begin{equation}
\hat{H}_{\rm int} = \hbar \times \bigg( 0.072 \frac{a}{a_0} -0.01 \bigg) {\rm Hz} ~\hat{J}_z^2.   
\end{equation}
It implies that the interaction within each double well of the superlattice potential can be exactly canceled when $\frac{a}{a_0} = 0.139.$

{\bf Multiparameter maximum likelihood estimation.}
We consider the set of $m$ joint measurement results $\vect{z}_{1} \equiv \{ z_1^{(1)}, ..., z_1^{(m)}\}$ and $\vect{z}_{2} \equiv \{ z_2^{(1)}, ..., z_2^{(m)}\}$ obtained for the two interferometers and introduce the rescaled variable $\tilde{z}_j^{(k)} = (z_j^{(k)}-C_j)/V_j$.
We assume that the output signal of each interferometer depends sinusoidally on the phase shift, with constant visibility $V_j$ and offset $C_j$. 
The offset is estimated as $C_j = \sum_{k=1}^m z_j^{(k)}$ and, for our sets of data, where $m \approx 30$, we find $C_j\approx 0$ (see e.g. caption of Fig.~\ref{fig 3}).
The visibility is estimated as $V_j = \max_k \vert z_j^{(k)}-C_j\vert$.
The phase difference $\Delta \phi$ and the uncorrelated noise $\sigma_{\Delta\phi}$ are estimated jointly, as the values that maximize the likelihood function $P(\vect{\tilde{z}}_{1}, \vect{\tilde{z}}_{2} \vert  \Delta \phi, \sigma_{\Delta\phi}) = \prod_{j=1}^m P(\tilde{z}_1,\tilde{z}_2\vert \Delta \phi, \sigma_{\Delta\phi})$.
A semi-analytical expression for $P(\tilde{z}_1,\tilde{z}_2\vert \Delta \phi, \sigma_{\Delta\phi})$ is obtained by assuming that the uncorrelated noise has a Gaussian distribution of width $\sigma_{\Delta\phi}$, giving
\begin{widetext}
    \begin{eqnarray}
P(\tilde{z}_{1}, \tilde{z}_{2} \vert  \Delta \phi, \sigma_{\Delta\phi}) = 
\frac{1}{(2 \pi)^{3/2} \sigma_{\Delta\phi} \sqrt{1-\tilde{z}_1^2} \sqrt{1-\tilde{z}_2^2}} 
\bigg( 
\sum_{k=0,1} e^{-\tfrac{[\theta + (-1)^k (\sin^{-1}(\tilde{z}_1) - \sin^{-1}(\tilde{z}_2))]^2}{2\sigma_{\Delta\phi}^2}} + e^{-\tfrac{[\theta + (-1)^k (\sin^{-1}(\tilde{z}_1) + \sin^{-1}(\tilde{z}_2)-\pi)]^2}{2\sigma_{\Delta\phi}^2}} 
\bigg). 
\end{eqnarray}
Further details about the multiparameter maximum likelihood method are reported in Ref.~\cite{ML}.
To associate error bars to the estimated values $\Delta \phi^{\rm est}$ and $\sigma_{\Delta\phi}^{\rm est}$ we use a Bootstrap method: we generate a large number of data $z_{1}$ and $z_{2}$ sampled according to $P(\tilde{z}_{1}, \tilde{z}_{2} \vert  \Delta \phi^{\rm est}, \sigma_{\Delta\phi}^{\rm est})$.
Data are divided in samples of $m$ values.
For every sample we obtain new estimated values of $\Delta \phi$ and $\sigma_{\Delta\phi}$. 
Finally, mean square fluctuations of the estimated values are computed. 

\end{widetext}


\bibliography{apssamp}

\begin{thebibliography}{65}
\bibitem[com()]{comment}
V$_j$ smaller than 1 is mainly due to unwanted excitations of the atoms to higher-lying states, finite temperature effects, and detection noise. Values of C$_j$ different from 0 are due to unwanted energy mismatch between the two wells during the splitting sequences.
\end{thebibliography}

\end{document}